\documentclass[12pt]{iopart}
\usepackage{iopams}
\usepackage{graphicx}
\usepackage{multirow}
\begin{document}
\title{He$_2^{3+}$ and HeH$^{2+}$ molecular ions in a strong magnetic field: the Lagrange mesh approach}
\author{Horacio Olivares Pil\'on}
\eads{\mailto{holivare@ulb.ac.be}}
\address{Physique Quantique, CP 165/82, Universit\'e Libre de Bruxelles, B 1050 Brussels, Belgium}
\date{today}
\begin{abstract}
Accurate calculations for the ground state of the molecular ions He$_2^{3+}$ and HeH$^{2+}$ placed in a strong magnetic field $B\gtrsim 10^{2}$~a.u. ($\approx 2.35 \times 10^{11}$G) using the Lagrange-mesh method are presented. The Born-Oppenheimer approximation of zero order (infinitely massive centers) and the parallel configuration (molecular axis parallel to the magnetic field) are considered. Total energies are found with 9-10 s.d. The obtained results show that the molecular ions He$_2^{3+}$ and HeH$^{2+}$ exist at $B > 100$\,a.u. and $B > 1000$\,a.u., respectively, as  predicted in \cite{Tu:2007} while a saddle point in the potential curve appears for the first time at $B \sim 80$~a.u. and $B \sim 740$~a.u., respectively.
\end{abstract}

\vspace{7cm}
\maketitle

\section{Introduction}

The neutron star surface is characterized by enormous magnetic fields, typically of
$B\sim 10^{12}\,$G ($\sim 10^3\,$a.u.)\footnote{$1$ a.u. $ \approx 2.35\times10^{9}$~G$ = 2.35\times10^5$~T.}. Even more, for some young neutron stars - magnetars, the magnetic field achieves values up to $B\sim 10^{16}\,$G ($\sim 10^7\,$a.u.). It raises a general question about the behavior of matter under such extreme magnetic fields (see for a review e.g. \cite{Lai:2001}). From a practical point of view, because of the intense magnetic fields, the content of a neutron star atmosphere became an issue.
What kind of chemical species are formed there? This question got major importance after 2002 when the Chandra X-ray observatory discovered in the spectra of radiation of the isolated neutron star 1E1207.4-5209 two  absorption features at $\sim 0.7$\,keV and $\sim 1.4$\,keV.

Extensive theoretical studies of traditional and exotic molecular systems in the presence of a strong magnetic field have been carried on (for a review see \cite{Tu:2007} and \cite{phrep}). It was found that for magnetic fields $B\leq 4.414\times10^{13}\,$G the most bound one-electron systems are the exotic molecular ions He$_2^{3+}$ and HeH$^{2+}$ besides the traditional molecular ion H$_2^+$. With the hypothesis of a neutron star atmosphere composed of mixed hydrogen-helium molecules, a model was proposed~\cite{Tur:2005} that could explain the two absorption features discovered by the Chandra X-ray observatory. Even more, if the magnetic field strength is increased, new exotic finite molecular chains start to be present. In particular, for $B = 2.1\times10^{4}$\,a.u. the potential curve of the exotic system Li$_2^{5+}$ start to develop a minimum~\cite{horop:2010}.

The aim of the present work is to make an accurate study of the two molecular ions 
${\rm He}_2^{3+}$ and  ${\rm HeH}^{2+}$ in the presence of a strong magnetic field $B \le 4.414\times 10^{13}$\,G where the relativistic corrections seem to be unimportant. The molecular axis is assumed to be aligned parallel to the magnetic field line because this configuration is optimal if a magnetic field is sufficiently strong \cite{phrep, BJS09}. The Born-Oppenheimer approximation is considered where the positively-charged centers are supposed to be infinitely massive.

Following~\cite{horop:2010} in the case of one-electron systems the {\it variational method}  and the {\it Lagrange-mesh method} are two methods which complement each other. The first one has been proven to be very efficient \cite{phrep} with physically motivated trial functions. On the other hand, the {\it Lagrange-mesh method} \cite{h2malla} nowadays provides the most accurate results for some simple one-electron molecular systems in the presence of a strong magnetic field.  In practice, the implementation of the Lagrange-mesh method calculations is made easier and often feasible with {\it a priori} knowledge of the equilibrium distance and the corresponding total energy. Such a knowledge is provided by a variational calculus. Thus, following the results presented in~\cite{Tu:2007} the {\it Lagrange-mesh method} is applied to check and further improve the variational results.

Atomic units $m_{e}= e = \hbar = 1$ are used throughout, although the energy is given in Rydbergs.

\section{One-electron molecular ion at the Born-Oppenheimer approximation}

\begin{figure}[!htp]
\centering
\includegraphics[height=5.0cm]{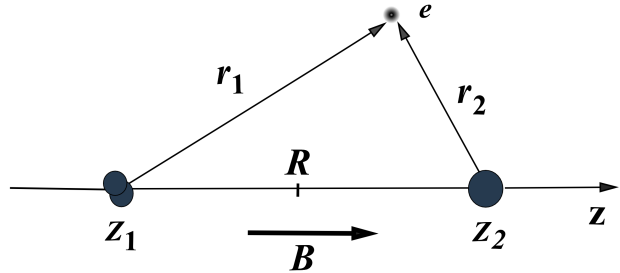}
\caption{Geometrical setting of the ground state $1\sigma_g$ for the helium containing molecular system of one-electron and two charged centers $Z_{1}=2$ and $Z_{2}=1,2$ placed in a magnetic field $B$  in parallel configuration along the $z$-axis. $R$ is the internuclear distance and $r_{1}$, $r_{2}$ are the distances between the electron and the charged centers $Z_{1}$ and $Z_{2}$, respectively.}
\label{fig:lix}
\end{figure}

Choosing the vector potential corresponding to a constant magnetic field ${\bf  B}=(0,0,B)$ in the symmetric gauge ${\cal A}= \frac{B}{2}(-y, \,x,\, 0)$, the Hamiltonian which describes two infinitely heavy centers of charges $Z_1=2$ and $Z_{2}=1,2$ situated along the $z$ axis and one electron placed in a uniform constant magnetic field directed along the $z-$axis, $\nobreak{{\bf B}=(0,0,B)}$ is given by
\begin{equation}
 \hat{\mathcal H} = -\Delta  -
\frac{2Z_1}{r_1} -\frac{2 Z_2}{r_2}\,+\frac{2\, Z_1Z_2}{R}
+ B {\hat L}_{z} + \frac{B^{2}}{4}\rho^{2} \,.
\label{ham-LiX}
\end{equation}
(see Fig.~\ref{fig:lix} for the geometrical setting). Here $\rho^{2}=x^{2}+y^{2}$ and $\hat L_{z}$ is the $z$-component of the electron orbital momentum, which is an integral of motion, $[{\hat H}, {\hat L}_z]=0$.

\section{The Lagrange-mesh method}

As it was pointed in~\cite{horop:2010} the Lagrange-mesh method and the variational method naturally complement each other. With this in mind, the Lagrange-mesh method \cite{h2malla,mesh,hmesh} is applied with the aim to confirm and improve the variational results presented in~\cite{Tu:2007}.
The most adequate coordinate system to describe a two-center molecular system of charges  $Z_{1}$ and $Z_{2}$ is the system of spheroidal coordinates $(\xi,\eta,\varphi)$ defined as
\begin{eqnarray}
\xi = \frac{r_{1}+r_{2}}{R} -1\,, \hspace{1cm} \eta= \frac{r_{1}-r_{2}}{R} \,,
\end{eqnarray}
(see Fig.\ref{fig:lix}) with $\xi\in(0,\infty)$, $\eta\in(-1,1)$ and the azimuthal
angle $\varphi \in (0,2\pi)$. Since the $z$-projection of the orbital angular momentum
$\hat L_z$ commutes with the Hamiltonian (\ref{ham-LiX}), it is replaced by its eigenvalue $m$ (magnetic quantum number). The wave function for the ground stated ($m=0$) should be nodeless and it can be written as~\cite{h2malla}
\begin{equation}
\Psi_{m=0}(\textbf{r}) = \frac{2}{(\pi R^{3})^{1/2}}\psi_{0}(\xi,\eta)\,.
\label{functot}
\end{equation}
The function $\psi_{0}(\xi,\eta)$  is expanded in the Lagrange basis as
\begin{equation}
\label{wavef}
\psi_{0}(\xi,\eta)=\sum_{i=1}^{N_{\xi}}\sum_{j=1}^{N_{\eta}}c_{ij}F_{ij}(\xi,\eta) \ ,
\end{equation}
where $N_{\xi}$ is the size in the $\xi$-direction and $N_{\eta}$ is in the $\eta$-direction. An explicit expression of $F_{ij}(\xi,\eta)$ and more details are given
in~\cite{h2malla}.

Taking the wave function (\ref{wavef}) and using the Gaussian quadratures associated to each coordinate, the Schr$\ddot{\mathrm o}$dinger equation for two-charged centers  gets the form of mesh equations (see \cite{h2malla})
\begin{equation}
\sum_{i'=1}^{N_{\xi}}\sum_{j'=1}^{N_{\eta}}[ T_{iji'j'}+V(hx_{i},\eta_{j})\,\delta_{ij}\delta_{i'j'} -E\, \delta_{ii'}\delta_{jj'} ]c_{i'j'}=0\, .
\label{matrixE}
\end{equation}
The kinetic energy matrix elements $T_{iji'j'}$ are calculated in \cite{h2malla}. The potential  $V(\xi, \eta)$ is evaluated at the zeros of the Laguerre $L_{N_{\xi}}(x_i)=0$ (scaled by the dimensionless parameter $h$) and the Legendre polynomials $P_{N_{\eta}}(\eta_j)=0$. Finally, the problem of solving the
Schr$\ddot{\mathrm o}$dinger equation is reduced to searching eigenvalues of the matrix equation (\ref{matrixE}).

\section{Results}

Results of the Lagrange-mesh calculations for the $m=0$ ground state of the molecular ions He$_2^{3+}$ and HeH$^{2+}$ placed in strong magnetic fields $B \geq 10^{2}\,$a.u.   in parallel configuration are presented in  Tables \ref{Thehe} and \ref{Theh}, respectively.
Both exotic systems, do not exist without the presence of a strong magnetic field. The results, demonstrate that the potential energy curve starts to display a well pronounced minimum at finite internuclear distance at the magnetic fields $B_{th} \sim 10^2$\, and $\sim 10^3\,$\,a.u., for the ions He$_2^{3+}$ and HeH$^{2+}$, respectively. It provides a theoretical indication about the possible existence of the bound exotic molecular ions ${\rm He}_2^{3+}$ and ${\rm HeH}^{+}$. The potential energy curve as a function of the internuclear distance $R$ is characterized by the presence of a potential barrier. At large internuclear distances the interaction of two charged centers becomes repulsive: the total energy curve approaches from above to the total energy of the atomic ion ${\rm He}^{+}$. The system is metastable towards the dissociation to ${\rm He}^{+} + p(\alpha)$ if the energy of ${\rm He}^{+}$ is lower than the minimum on the potential curve. Both considered systems display two characteristic properties of the Coulomb systems in a magnetic field: as the strength of the magnetic  field increases, they become more bound (the binding energy grows) (i)  and more compact (the equilibrium distance decreases) (ii).

Tables \ref{Thehe} and \ref{Theh} present the  equilibrium internuclear distance $R_{eq}$, the corresponding  total  $E_{t}^{min}$  and binding  $E_{b}= B - E_{t}^{min}$ energies, the He$^{+}$ total energy from which the dissociation energy $E_{diss}=E_{t}^{min} - E_{t}^{{\rm He}^{+}}$ can be obtained, as well as  the position $R_{max}$ of the maximum $E_{t}^{max}$ of the barrier  and its height  {\it i.e.} the difference $\Delta E =E_{t}^{max}-E_{t}^{min}$ for the molecular ions ${\rm He}_2^{3+}$  and  ${\rm HeH}^{2+}$, respectively.

The lowest vibrational energy $E_{0}^{vib}$ is calculated by using the harmonic oscillator approximation around the equilibrium position. It allows us to define the zero point energy and is presented in Tables \ref{Thehe} and \ref{Theh}.
The computer code JADAMILU~\cite{JDM} was used for the lowest-eigenvalue search in the mesh calculations.

\subsection{The molecular ion ${\rm He}_{2}^{3+}$}
\begin{table}[!htbp]
\caption{\label{Thehe} Ground state $1\sigma_g$ of the molecular ion ${\rm He}_{2}^{3+}$ in a magnetic field $B$: equilibrium internuclear distance $R_{eq}$, total $E_{t}^{min}$ and binding $E_{b} = B-E_{t}^{min}$ energies; He$^{+}$ and H$_2^+$ binding energies; $R_{max}$ of the maximum at the potential curve; height of the barrier $\Delta E=E_{t}^{max}-E_{t}^{min}$ and lowest vibrational energy $E_0^{vib}$. For each value of the magnetic field the second line presents the variational results from \cite{Tu:2007}}
\lineup
\resizebox{16cm}{!}{
\begin{tabular}{cllllllllc}
\br
$B$&$R_{eq}$&$E_t^{min}$[Ry]&$E_b$[Ry]&$E_b^{{\rm He}^+}$[Ry]&$E_b^{{\rm H}_2^+}$[Ry]&$R_{max}$&$\Delta E$[Ry]&$E^{vib}_0$[Ry]&\\
\mr
\multirow{2}{*}{10$^2$\,a.u.}
&0.77926&83.47593826&16.52406174&19.12109305&10.292162923&1.0219&0.0331&0.026&Present\\
&0.780  &83.484     &16.516     &19.11      &10.291&1.02 &0.033 &0.026&\cite{Tu:2007}\\
\bs
\multirow{2}{*}{10$^{12}$\,G}
&0.42004&396.8490192&28.6828956 &30.93148909&17.1522314 &0.9044&1.0250&0.100&Present\\
&0.420  &396.864    &28.668     &30.87      &17.143&0.90 &1.024 &0.100&\cite{Tu:2007}\\
\bs
\multirow{2}{*}{10$^3$\,a.u.}
&0.30923&960.7139916&39.2860084 &40.5413906&22.7816779  &0.8140&2.4633&0.170&Present\\
&0.309  &960.732    &39.268     &40.40     &22.779&0.82&2.466 &0.169&\cite{Tu:2007}\\
\bs
\multirow{2}{*}{10$^{13}$G}
&0.19314&4190.162674&65.156475  &62.392739&35.760513 &0.6818&7.3154&0.369&Present\\
&0.193  &4190.182   &65.137     &61.99    &35.754&0.70 &7.328 &0.366&\cite{Tu:2007}\\
\bs
\multirow{2}{*}{10$^4$\,a.u.}
&0.14955&9913.740241&86.259759  &79.097975&45.80516 &0.6175&12.2422&0.544&Present\\
&0.150  &9913.767   &86.233     &78.43    &45.797&0.62&12.25&0.561&\cite{Tu:2007}\\
\bs
\multirow{2}{*}{4.414$\times10^{13}$G}
&0.1256 &18677.83406&105.14466&93.45755&54.54632&0.576&17.1720&0.732&Present\\
&0.126  &18677.857  &105.121  &92.53&54.502  &0.58 &17.19  &0.739&\cite{Tu:2007}\\
\bs
10$^5$\,a.u.&0.08174&99828.5786 &171.4214   && --& -- & --&1.451&Present\\
\br
\end{tabular}}
\end{table}

The results for the ground state $1\sigma_g$ of the  symmetrical system He$_2^{3+}$ for magnetic fields $B\ge 10^{2}$\ a.u. in parallel configuration are presented in Table~\ref{Thehe}.
For each value of the magnetic field shown in Table~\ref{Thehe}, the results are compared with those presented in Turbiner and L\'opez Vieyra \cite{Tu:2007} in the frame of the variational method. In all cases the Lagrange-mesh total energies are essentially more accurate than the corresponding variational results. For the binding energies, the relative accuracy is $\sim 0.05\%$ at $B=10^{2}$\ a.u. while it is $\sim 0.02\%$ at $B=4.414\times10^{13}$\ G.

Increasing the magnetic field value from zero, the potential curve first does not present an indication of any bound state. However, for $B \sim 80$~a.u. a saddle point appears and eventually for a magnetic field $B \sim 10^{2}$\ a.u. the potential energy curve starts to display a well pronounced minimum for $R\sim0.78$\ a.u. indicating the existence of a metastable state of the molecular ion ${\rm He}_{2}^{3+}$, unstable towards decay to ${\rm He}_{2}^{3+} \rightarrow {\rm He}^{+} +\, \alpha$. For this value of the magnetic field the energy curve shows that the potential barrier is high enough to keep at least one vibrational level, $E_0^{vib} < \Delta E$. Hence, the system is stable against vibrations. The height of the barrier increases when the magnetic field grows and the system becomes more stable allowing more vibrational modes. Both $\Delta E$ and $E_0^{vib}$ increase monotonously with the magnetic field, their ratio   $\Delta E/E_0^{vib}$ increases from $\sim 1$ for $B = 10^2$\ a.u. up to $\sim 23$ for $B=4.414\times10^{13}$\,G. At the same time, the dissociation energy $E_{diss}=E_t^{min}-E_t^{\rm He^+}$ decreases and eventually, for magnetic fields $B \sim 10^{13}$\ G., the total energy becomes smaller than the total energy of the ${\rm He}^+$ atomic ion. Thus, the system becomes stable towards the decay ${\rm He}_{2}^{3+} \rightarrow {\rm He}^+ +\alpha$. The qualitative behavior of this molecular system is typical: when we increase the magnetic field strength, the molecular ion ${\rm He}_{2}^{3+}$ becomes more compact (the equilibrium distance decreases) and more bound (the binding energy increases), see e.g.[3].

\subsection{The molecular ion ${\rm HeH}^{2+}$}

\begin{table}[!htp]
\caption{\label{Theh} Ground state $1\sigma_g$ of the molecular ion ${\rm HeH}^{2+}$ in a magnetic field $B$: equilibrium internuclear distance $R_{eq}$, total $E_{t}^{min}$ and binding $E_{b} = B-E_{t}^{min}$ energies; He$^{+}$ and H$_2^+$ binding energies; $R_{max}$ of the maximum at the potential curve; height of the barrier $\Delta E=E_{t}^{max}-E_{t}^{min}$ and lowest vibrational energy $E_0^{vib}$. For each value of the magnetic field the second line presents the variational results from \cite{Tu:2007}}
\lineup
\resizebox{16cm}{!}{
\begin{tabular}{cllllllllc}
\br
$B$&$R_{eq}$&$E_t^{min}$[Ry]&$E_b$[Ry]&$E_b^{{\rm He}^+}$[Ry]&$E_b^{{\rm H}_2^+}$[Ry]&$R_{max}$&$\Delta E$[Ry]&$E^{vib}_0$[Ry]&\\
\mr
\multirow{2}{*}{10$^{3}$\,a.u.}
&0.3161 & 962.6035199&37.3964801&40.5413906&22.7816779&0.4215&0.0581664&0.126&Present\\
&0.316  & 962.635    &37.365    &40.40     &22.779    &0.424 &0.061    &0.13 &\cite{Tu:2007}\\
\bs
\multirow{2}{*}{10$^{13}$G}
&0.1846 & 4195.651303&59.667846 &62.392739&35.760513 &0.3756&1.189682 &0.413&Present\\
&0.185  & 4195.693   &59.626    &61.99    &35.754    &0.375 &1.197    &0.41 &\cite{Tu:2007}\\
\bs
\multirow{2}{*}{10$^{4}$\,a.u.}
&0.14073& 9922.639585&77.360415 &79.09797 &45.80516  5&0.3416&2.65622  &0.604&Present\\
&0.142  & 9922.697   &77.303    &78.43    &45.797    &0.341 &2.651    &0.65 &\cite{Tu:2007}\\
\bs
\multirow{2}{*}{4.414$\times10^{13}$G}
&0.1190 & 18690.07669&92.90203  &93.45755 &54.54632  &0.3185&4.21166  &0.878&Present\\
&0.119  & 18690.121  &92.858    &92.53    &54.502    &0.318 &4.22     &0.88 &\cite{Tu:2007}\\
\br
\end{tabular}
}
\end{table}

Table~\ref{Theh} presents the results for the $1\sigma$ state of the asymmetric two-center system ${\rm HeH}^{2+}$  for magnetic fields $B \geq 10^{3}$\ a.u.  in parallel configuration.
For each value of the magnetic field shown in Table~\ref{Theh}, the lower line corresponds to the variational results given in~\cite{Tu:2007}. Systematically, the Lagrange-mesh results give much more accurate results for all studied magnetic fields. For the binding energy, the relative improvement of the mesh calculations is $\sim 0.08\%$, for $B= 10^{3}$\ a.u. up to  $\sim 0.05\%$ for $B=4.414\times10^{13}$\,G.

As a function of the magnetic field a saddle point appears on the potential curve when the magnetic field value is $B\sim 740$~a.u. Increasing the magnetic field until $B\sim 10^3$\ a.u., a well pronounced minimum on the potential curve emerges at $R\sim 0.32$\ a.u.,  indicating the formation of a metastable state, unstable for the decay to ${\rm HeH}^{2+} \rightarrow {\rm He}^{+} + p$. The potential barrier is small and it does not allow to keep a single vibrational level $E_0^{vib} > \Delta E$. Finally, when the magnetic field strength becomes $B \sim 10^{13}$\,G, the potential energy well is sufficiently deep to keep more than one vibrational state. Also, the height of the barrier increases when the magnetic field is increased and the system becomes more stable against vibrations. The ratio $\Delta E/E_0^{vib}$ increases from $\sim 3$ for $B=10^{13}$\,G up to $\sim 5$ for $B=4.414\times10^{13}$\,G. The dissociation energy $E_{diss}=E_{t}^{min}-E_{t}^{\rm He^+}$ decreases but the system remains unstable towards the decay ${\rm HeH}^{2+}   \rightarrow {\rm He}^+ + p$ for all magnetic fields studied. The system is always stable towards the decay ${\rm HeH}^{2+} \rightarrow {\rm H} + \alpha$ due to the fact that the hydrogen atom in a magnetic field is the least bound  one-electron system.

\section{Conclusions}

It is presented a study in the non-relativistic frame and the Born-Oppenheimer approximation of zero order of the ground state of the molecular ions ${\rm He}_2^{3+}$ and ${\rm HeH}^{2+}$ in the presence of a strong magnetic field $B\le 4.414\times10^{13}$~G in parallel configuration. Although  the {\it Lagrange-mesh method} is applied, the {\it variational} results presented in~\cite{Tu:2007} are taken into account as a reference point for the mesh calculations (see~\cite{horop:2010}).

The  results give clear indications that the exotic molecular ions
${\rm He}_2^{3+}$ and ${\rm HeH}^{2+}$ begin to exist as metastable states starting at the threshold magnetic fields $B_{th} \sim 10^2$ and  $\sim 10^3$\ a.u., respectively.  As the magnetic field increases the potential wells of both systems become sufficiently deep to keep more than one vibrational state.
Eventually, for $B= 10^{13}$\,G the ion  ${\rm He}_{2}^{3+}$ becomes stable. The ${\rm HeH}^{2+}$ ion remains unstable towards decay ${\rm HeH}^{2+}\rightarrow{\rm He}^{+} + p$ in the whole domain of magnetic fields considered $B\leq 4.414\times10^{13}$\,G.
For both systems and all studied magnetic fields the obtained results are essentially more accurate then those presented in~\cite{Tu:2007}.

These molecular ions have a similar behavior: when the magnetic field increases, each molecular ion becomes more compact and more bound. For all magnetic fields  the most bound helium-containing molecular system is ${\rm He}_{2}^{3+}$, $i.e.$, $E_{b}^{{\rm He}_{2}^{3+}} > E_{b}^{{\rm HeH}^{2+}}$. From Tables \ref{Thehe} and \ref{Theh} it can be seen that the H$_2^+$ molecular system  placed in a magnetic field is the less bound two-center one-electron molecular system at $B \geq 10^{13}$\,G.

\ack{The author wish to thank A.V. Turbiner for the interest to work, valuable discussions and a careful reading of the manuscript and D. Baye for reading the manuscript and for comments. The present work is supported in part by CONACyT (M\'exico) through a postdoctoral grant.}

\end{document}